DEGREE PROJECT

# Finding differences in perspectives between designers and engineers to develop trustworthy AI for autonomous cars

Karl Rikard Larsson
Gustav Jönelid

[This page intentionally left blank]

## Abstract

In the context of designing and implementing ethical Artificial Intelligence (AI), varying perspectives exist regarding developing trustworthy AI for autonomous cars. This study sheds light on the differences in perspectives and provides recommendations to minimize such divergences. By exploring the diverse viewpoints, we identify key factors contributing to the differences and propose strategies to bridge the gaps. This study goes beyond the trolley problem to visualize the complex challenges of trustworthy and ethical AI. Three pillars of trustworthy AI have been defined: transparency, reliability, and safety. This research contributes to the field of trustworthy AI for autonomous cars, providing practical recommendations to enhance the development of AI systems that prioritize both technological advancement and ethical principles.

# Acknowledgments

We would like to take this opportunity to express our gratitude to Hamam Mokayed and Johan Wenngren for their contributions and support throughout our thesis project.

To Hamam Mokayed at Luleå University of Technology, we are immensely grateful for your invaluable guidance and assistance right from the inception of this project. Your expertise, dedication, and insights have played a pivotal role in shaping the direction of our research and elevating the overall quality of our work. Your assistance in establishing crucial contacts, providing access to publications, and your continuous availability for discussions and feedback have enriched our learning experience and made this journey all the more meaningful.

We would also like to extend our deepest appreciation to Johan Wenngren at Luleå University of Technology for his role as our supervisor. Your guidance, and constructive feedback have been instrumental in shaping our research methodology and ensuring the academic rigor of our project. Your support and dedication have served as a constant source of inspiration throughout every stage of this process.

Furthermore, we would like to extend our thanks to all the participants and individuals who shared their time, insights, and expertise with us. Your cooperation, willingness to contribute, and openness in sharing your experiences have significantly enriched our research, making it a more comprehensive and insightful endeavor.

**Table of Contents**

# 1.0 Introduction

The unprecedented rate of technological development in recent times has given rise to Industry 4.0, which is characterized by its impact on profits and the integration of cutting-edge technologies (Leng et al., 2022; Xu et al., 2021). Despite its many advantages, the fast-paced industrialization associated with Industry 4.0 has had negative consequences for both the environment and the safety of workers. These include increased energy consumption and greenhouse gas emissions, automation-related job losses and workplace accidents and more. This has prompted the need for a more sustainable Industry 4.0 framework. Industry 4.0 offers a profound insight into how advanced technology is reshaping the digital ecosystem of our present society. It provides diverse perspectives on the role of machines in replacing human attributes within both the manufacturing and developmental domains. The current shift towards Industry 5.0 aims to address these consequences by integrating human technicians into the manufacturing process to increase efficiency and promote human-autonomous machine collaboration (Javed et al., 2023; Leng et al., 2022; Xu et al., 2021). Industry 5.0 is a concept that aims to promote value-driven automation in the smart industry sector in Europe. At its core are the values of resilience, sustainability, and human-centricity. Industry 5.0 aims to harness the benefits of new technology while also acknowledging the current state of existing jobs and the limitations of the world, with a focus on prioritizing workers' well-being (Breque et al., 2021).

As the field of Industry 5.0 and beyond continues to evolve, the interplay between ethical considerations and technological feasibility is becoming increasingly important. In parallel, the artificial intelligence (AI) sector is also undergoing a transformation with similar propositions as the industri 5.0, with a growing emphasis on human-centered AI (HCAI) and ethical considerations (Javed et al., 2023; Auernhammer, 2020; Shneiderman, 2020a; Shneiderman, 2020b;). This interplay between industri 5.0 and AI enables for human ethical consideration both in the development processes of AI systems, as well as the manufacturing of the car itself. In addition, this collaboration allows AI to augment human drivers, making transportation safe and more efficient. Including industry 5.0 in the discussion sheds light on the integration of AI technologies in the automotive sector and underscores the importance of human skills alongside autonomous systems. This focus is critical to ensuring that new technologies are developed and implemented in a responsible and sustainable manner. In addition, by providing users with a fair degree of human control over an AI system, it is hoped that the system will be more reliable, safe, and trustworthy (Shneiderman, 2020b).

AI is widely implemented and utilized across various sectors, including the field of transportation, specifically in vehicle intelligence systems (VIS). The integration of AI in VIS is crucial for enhancing the performance and functionality of contemporary vehicles. AI plays a pivotal role in improving these systems by enabling vehicles to analyze complex data and understand natural scene images (Khan et al., 2022; Mokayed et al., 2016), learn from patterns (Mokayed et al., 2023), and make intelligent decisions in real-time (Quan et al., 2022). By incorporating AI into the vehicle intelligence system, we can unlock fresh possibilities for autonomous driving, advanced driver-assistance features, and personalized driving experiences.

However, there are differing views on the nature of AI and its role in society. Auernhammer (2020) argues that there are two different perspectives on AI: the rationalistic view and the design view. The rationalistic view focuses on advancing computer systems that mimic human intelligence, while the design view sees AI as a tool to improve human conditions and emphasizes the interaction between humans and computers.

**Problem statement**

While there is a growing emphasis on incorporating human-centricity and value-driven automation in these fields, there are differing views on the role of AI in society and how it should be developed and implemented. Therefore, there is a need for further research and discussion on how to ensure that these views are aligned and that the technology is developed and used in a responsible and sustainable manner, while also prioritizing the well-being of workers and the environment.

**1.1 Purpose**

In light of these differing perspectives, this study investigates the differences between designers and engineers with regard to ethics when developing autonomous vehicles in a team setting. This study proposes a framework that considers the ethical implications of autonomous systems and supports ethical decision-making in multi-disciplinary teams. However, it's important to emphasize that these solutions must align with the ethical framework and values and be carefully designed with human well-being in mind, given the potential impact on society.

The purpose of this paper is not only to provide insights but also to highlight the differences in perspectives regarding ethical considerations related to autonomous vehicles. By presenting a brief outline of the issues at hand, we aim to facilitate discussions and foster collaboration towards the development of a holistic and ethical framework for autonomous cars. In addition to characterizing the differences, we will provide a detailed description of each difference and offer recommendations that can be used as strategies on how to reduce these divergences.

This study specifically focuses on the ethical considerations involved in the development of AI for autonomous cars. It is worth noting that once these cars are manufactured and on the road, the driver also has ethical responsibilities. However, our research is primarily focused on the development phase, and aims to provide insights into how developers can ensure that their AI systems are developed in an ethical and trustworthy manner.

In addition to addressing practical concerns, our goal is to contribute to the broader field of artificial intelligence by highlighting the need for a multidisciplinary approach. Therefore, we encourage those involved in the development process of autonomous cars to consider the insights presented in this paper and to engage in collaborative efforts towards a more ethical and human-centered approach. Specifically, we aim to answer these questions:

*What is the difference in perspective between designers and engineers with regard to ethics when developing autonomous vehicles?*

*What strategies can be implemented to bridge the gap in perspectives between designers and engineers regarding the development of autonomous cars?*

To avoid confusion, it is crucial to define the terms *engineer* and *designer* in the context of our study. According to Charisi et al., (2017), intelligent autonomous systems involve three parties: designers (including developers and engineers), end-users (owners or customers), and various government and trade regulatory bodies and insurance agents. Our study used these parties as a guideline but delved deeper into the designer category by distinguishing between designers and engineers of AI systems. Specifically, we use the term engineer to refer to individuals who specialize in back-end development, such as AI coders or researchers focused on the technical aspects of AI. In contrast, designers are more oriented towards front-end development and user-centered design in the context of UX-researcher, UX/UI-designer, while researchers within these fields are considered professors specialized in system science and informatics.

## 2.0 Theory

### 2.1 Trust & disrupting the business model

The traditional business model of selling cars is being disrupted by new mobility solutions. One of the most innovative companies in this space is Tesla Motors, which offers supplementary services through a modular approach. For instance, the Model S, Tesla's flagship car, is designed as a platform that can accommodate additional services. These include autonomous driving capabilities that can be uploaded and updated over the air for a fee. To make this possible, the company must develop cars with the services in mind and manage the development process accordingly (Mahut et al., 2017). In short, this means that car manufacturers are moving towards selling cars as a service, with smart features instead of selling just a vehicle with the only purpose to transport the driver from A to B.

A survey conducted in Germany 2015 revealed that user acceptance of autonomous driving is low. Nearly half of the 1003 car buyers surveyed were skeptical, with only 5% preferring fully autonomous driving, approximately 49% preferred traditional driving and 43% preferred semi-autonomous driving (Hengstler et al., 2016).

Building trust in autonomous driving is crucial for reducing perceived risk and encouraging the adoption of such technologies. At its core, trust is a fragile and evolving phenomenon that requires the willingness to be vulnerable to the actions of another. As automation technology continues to advance, trust evolves across different dimensions, including predictability, dependability, and faith (Hengstler et al,. 2016). According to Hengstler et al. (2016), trust initially stems from predictability, where users develop trust based on the system's consistent and expected behavior. As the system demonstrates dependability and reliability over time, trust deepens. Ultimately, faith emerges when users rely on the technology with confidence, knowing that it will perform as expected.

In the case of radically new technologies like autonomous driving, establishing initial trust is crucial in overcoming perceived risks. Perceived risk is defined as the level of uncertainty surrounding the potential failure of a new product or the probability of the product not functioning as intended. In the context of AI, perceived risk is related to the fact that control is delegated to a machine, making the driver feel uneasy about allowing a machine to make decisions and take actions on its own, without human intervention (Hegstler et al., 2016).

### 2.2 Human involvement

In an article from Hjetland (2015) she describes former research where the developers/ engineers have concluded that the human is the weakest link in the ecosystem. This is mainly in the order where the machine or in this case the vehicle can perform tasks more precise than humans can. Because the machine doesn't have human attributes (hunger, tiredness, etc.) which makes it more efficient and reliable. But there is a possibility that the machine breaks down or fails which requires human interaction (Hjetland, 2015). With the possibility of failure it reduces the trust from humans to the machine which can lower the adoption rate. Hjetland (2015) also presents the importance of the trustworthiness towards an autonomous system, it could be misused in the course of human interaction. With the underestimated- or overestimated trust towards an autonomous system it can lead to accidents where human interaction is or isn't required. This clarifies the importance of trust and safety when human interaction is involved with an autonomous system. Hjetland (2015) also examined various articles and found that one of the major concerns with autonomous vehicles is the level of trust that people have in them, which aligns with the results from the survey conducted in Germany where only 5% prefer fully autonomous driving (Hengstler et al., 2016).

In contrast to viewing humans as the weakest link in the ecosystem, Shneiderman (2020b) presents that both the human and the machine are valuable in different fields. The machine is more valuable when it is reliable and the human when the system requires human control, which machines regularly do. If the human is the weakest link in the ecosystem then the systems should not only be viable when it's reliable, but instead all the time.

Aurenhammer (2020) highlights the importance of involving humans in the development of AI, to make it human centric. To involve humans in the development it requires a human-centered approach in both the philosophy and the designing of AI. The article presents two different views on AI which is rationalistic and design, the main differences between them is that one is more user-centric while the other one focuses more on the achievements of the machine. Aurenhammer (2020) describes that human centered design (HCD) creates an opportunity for humanity to see the benefits of involving a human-centric design approach when developing AI, to create a society that originated from human needs.

## 2.3 Artificial intelligence

The idea of artificial entities, robots or creatures with their own intelligence is ancient and has existed as a human storytelling tradition for millennia (Wärnestål, 2021). Technical advances in algorithm research, the availability and large amounts of data, the internet as infrastructure, increased processing power and the spread of mobile devices have all contributed to the AI revolution that we are in the midst of in the early 2020s (Wärnestål, 2021).

AI is a technology that can learn and make decisions in complex situations, even interact with people and maintain social relationships. This kind of intelligence mirrors the decision-making process observed in humans and some animals. Yampolskiy. (2020) defined AI as a system that can adjust to its surroundings and make decisions even if it doesn't have all the information it needs. This definition is helpful when we think about how autonomous cars work, because they need to be able to navigate different environments without a lot of training or knowledge.

There are two main types of AI: narrow AI (also known as weak AI) and general AI (also known as strong AI). Narrow AI refers to AI systems that are designed to perform a specific task or a narrow range of tasks. These systems are usually based on data-driven machine learning algorithms and can perform their task with high accuracy and efficiency (Wärnestål, 2021). In contrast, general AI refers to AI systems that exhibit intelligence across a wide range of tasks, similar to human intelligence (Wärnestål, 2021).

With all this in mind, we considered AI and autonomous cars as the following when conducting this study: "Machines that can perform tasks that typically require human intelligence. Autonomous cars are an example of AI that use sensors and machine learning to navigate roads safely, and can learn from previous experiences to improve their driving capabilities."

Within the field of AI, responsible AI is a rising research topic that aims to ensure safe and ethical use of artificial intelligence. Responsible AI includes a set of principles, such as, but not limited to explainable AI (XAI), which focuses on promoting transparency, fairness, safety, security, integrity, accountability, and ethics. Its primary goal is to generate more explainable models while retaining a high level of learning performance, allowing individuals to comprehend and trust the decisions made by AI. (Barredo Arrieta et al., 2020)

The distinction between the transparency and explainability needs of end users and developers is crucial. Developers require in-depth understanding of the underlying attributes and algorithms, while end users need more comprehensive and simplified explanations about the system itself. To develop systems that satisfy both groups is a big challenge, and most interfaces tend to focus on either developers or end users (Ozmen et al., 2023).

There is a high need for explainability based on interpretable transparency in case an AI system fails or makes ethically relevant decisions. The individual affected would want to ensure that mistakes that were made are not repeated and that distribution of outcomes are based on fair, legal, and ethical decision-making mechanisms in society. The importance of explainability becomes even more pronounced in light of possible responsibility gaps that may arise (Ozmen et al., 2023).

## 2.4 Definition for autonomous cars

Autonomous driving is at the moment one of the most intensively researched technologies in the transportation domain (Hengstler et al., 2016). To the layperson, the definition of autonomous vehicles may appear self-evident, characterized simply as a vehicle capable of self-driving. But going in-depth to its actual meaning, it refers to that the vehicles are equipped with Advanced Driver Assistance Systems (ADAS) that help the driver make decisions and maneuvers in day-to-day driving as well in critical situations like a collision (Rödel et al., 2014). This definition might seem quite broad, as there are different levels of autonomy. Therefore, an automotive standardization for autonomous cars was defined by the Society of Automotive Engineers (SAE). These levels range from no autonomy (Level 0) to full autonomy (Level 5), and each level represents an increasing degree of automation and decreasing reliance on human intervention (Ondruš et al., 2020).

*Level 0: No automation:* The driver is in control of the vehicle at all times.

*Level 1: Driver assistance*: The driver is still in control, but the car can assist the driver with steering or braking.

*Level 2: Partial automation:* The car can control both steering and acceleration/deceleration, but the driver is still responsible for monitoring the environment and taking over control when necessary.

*Level 3: Conditional automation:* The car can handle most driving tasks under certain conditions, but the driver must be ready to take control when the system requests it.

*Level 4: High automation:* The car can handle most driving tasks without human intervention, but only under specific conditions or in specific environments.

*Level 5: Full automation:* The car can handle all driving tasks in all conditions and environments, and no human intervention is required.

## 2.5 Ethics Matters for Autonomous Cars

The field of ethics has been studied from various perspectives where all focused around logical reasoning and ethical evaluations. In other words, ethics explains the differences in right or wrong from the humans actions, what is morally right to do. It is concerned with determining what is right or wrong, good or bad, and just or unjust in human behavior and decision-making (Chonko, 2012). However, as we shift from human decision-making to system-based decision-making, it introduces an additional layer of unexpected behaviors, dilemmas and scenarios that increase the complexity.

A study by Massachusetts Institute of Technology (MIT) on the "MIT Moral Machine" offers valuable insights into the influence of cultural and societal factors on moral decision-making in autonomous vehicles, especially in death-threatening situations (Wärnestål, 2021). Results of the study indicated that individuals from Southeast Asian countries were more prone to prioritize the well-being of an older individual over a younger person in a moral dilemma, as compared to those from Western cultures. Conversely, individuals from Hispanic and French cultures were more likely to spare an athletic individual over someone overweight. In regions characterized by high economic inequalities, such as Colombia, participants tended to prioritize the well-being of a businessman over a homeless individual (Wärnestål, 2021). Having this context in mind, what would be the appropriate course of action? Let us imagine a scenario where you're driving an autonomous car and your car is faced with a terrible choice: it must either swerve left and hit an eight-year-old girl or swerve right and hit an 80-year-old grandmother. Due to the current speed of the car, either victim would most likely not survive the impact. If the car does not swerve, both individuals will be hit and most likely killed, making it imperative to swerve one way or the other (Maurer et al., 2016). The question arises, what is the ethically correct decision in such a scenario? If you were responsible for programming the self-driving car, how would you program it to behave? It may seem tempting, based on the previous research from MIT, to program the car to hit the younger person in East Asian countries and the older person in Western countries, but this approach is ethically questionable.

Moving away from the ethical perspective and into the perspective of the AI, let us imagine an autonomous car facing an imminent crash where it must make a terrible choice: it must either swerve a motorcyclist wearing a helmet or a motorcyclist without one. The safety of the car and its occupants may not be significantly affected by whether the motorcyclist is wearing a helmet or not, but the impact of a helmet hitting the car window may cause more damage to the car. Therefore, we must consider that if we optimize the algorithm without ethical considerations, the program may become selfish and choose to swerve towards the person without a helmet to reduce the damage done to itself (Maurer et al., 2016). It is crucial to acknowledge that the presence of a helmet is of importance to the safety of the motorcyclist. In this dreadful scenario, it seems reasonable to program a good autonomous car to prioritize the safety of the motorcyclist and swerve into the one wearing the helmet.

These scenarios and findings underline the substantial impact that cultural, environmental and context differences have on ethical decision-making in autonomous vehicles. It can be argued that these differences and development in regards to these concerns may be one of the most important factors of ethics, as they provide answers to the question of "whose ethical behavior?", as ethical behavior is different to different cultures, people and may differ in different contexts and raise important considerations regarding the adaptability of autonomous cars to cultural variations when operating in different countries (Wärnestål, 2021). Hence, the examination of cultural and societal factors in ethical decision-making is crucial in the development of ethics for autonomous vehicles and opens up hard questions such as "how should a self-driving car actually behave in traffic, and should it change as soon as you want to drive abroad where cultural perceptions may look different?" (Wärnestål, 2021).

It is important to acknowledge the complexities and challenges posed by the ethical considerations surrounding autonomous vehicles. However, it should be noted that the scope of this thesis is limited to the knowledge and ethical perspectives of researchers and workers within the European Union. Reasoning for this limitation is that the study takes place at a university based in Sweden and the collected data from interviews will take place in Sweden. In addition to this, we have also narrowed down the scope to not include technical or in-depth algorithms or machine learning techniques, only ethical aspects for the development of autonomous cars.

## 2.6 Transparency, reliability and safety

Transparency, reliability, and safety are identified as the key pillars for achieving trustworthy AI, as they contribute to creating systems that can be trusted by users, stakeholders, and society as a whole. Transparency allows users to understand how an AI system arrived at its decision, thus building trust with its users and stakeholders (Barredo Arrieta et al., 2020). Reliability, on the other hand, ensures that the system performs consistently and predictably, emphasizing human accountability, fairness, and comprehensibility (Shneiderman., 2020b).

Safety is also critical, as it ensures that AI systems behave as intended and do not cause unintended harm or unexpected consequences (Mikalef et al., 2022).

In more detail, a transparent system enables humans to see and understand the processes and actions taken by a machine. Barredo Arrieta et al. (2020) describes transparency as the opposite of a black box, to understand the machine itself. Which aligns with the findings of Charisi et al., (2017) where they argue that technology should not be a black box. The technology should instead be open for analysis and maintenance, in order for end users to interact with the technology in a safe manner. Transparency is an increasing subject in the development of systems, because the machines that are employed today make critical prediction decisions. Barredo Arrieta et al. (2020) points out that humans do not adopt new technologies that are not deemed trustworthy, which increases the importance of implementation of transparency in systems. With a transparent system, we have the opportunity to create credibility to the system and its processes, and gain trust to its users.

To achieve reliability in systems there need to be technical methods for human responsibility that can be achieved from audit trails and analytical techniques in order to review failures, benchmarkings for validation and verification, continuous review of collected data, design strategies for different stakeholders to build confidence (Shneiderman., 2020b). Shneiderman (2020b) proceeds to explain that a reliable system requires studying past data to see what failed and make the system reliable for users. In simple words, a reliable autonomous car is one that can operate safely and efficiently under a wide range of conditions, without breaking down or malfunctioning. This is particularly important in autonomous cars as any failure or malfunction in these systems could have serious consequences, potentially resulting in accidents or other safety hazards.

In the development of AI, safety is one of the most important factors for it to succeed. As the development of AI continues, it brings new ethical questions that need to be addressed and transformed into necessary requirements for the development of AI applications. Mikalef et al (2022) highlight the crucial role of safety and security in the design and implementation of AI applications, which has its roots for a trustworthy system. Shneiderman (2020b) presents safety as a key component to create a high performance system, and reveals that there are four methods to reach cultural safety. The four methods are that (1) the leadership should be committed to safety, (2) the reporting of failures should be open, (3) internal oversight for problems and the plans for the future, (4) and involvement of public reports. With an improved view of management strategies we can further improve the safety when implementing AI systems (Shneiderman 2020b) In simple words and summarization of safety;  a safe system is one that has been designed to prevent as much unintended harm and unexpected consequences as possible (Mikalef et al., 2022).

## 3.0 Methodology

### 3.1 Research approach

The problem area has limited research where there is no clear evidence that there is a difference in perspectives related to development of autonomous cars. However, there are studies acknowledging the shift in trend for AI, for example shift towards HCAI. This shift will cause divergences in perspectives due to adding more dimensions of complexity to the moral questions. Thus, the area that shows the direct difference does not exist and that is why we have chosen to illuminate this area with an explorative study. An explorative study is a research design that is used to investigate a question or problem that has not been clearly defined, and is commonly used when the topic is new or there is a limited existing knowledge. Explorative studies are designed to generate hypotheses, ideas and insights that can be used to guide further research and are mostly used to gain a better understanding of the research problem (Jacobsen, 2017).

It needs to be clear that ethics is a complex concept that is difficult to quantify. Therefore, using a qualitative approach has provided a profound awareness of the problem area and phenomenon that is being investigated (Jacobsen, 2017). Given our qualitative approach, a case study is an appropriate research strategy to employ in this research. A case study is well-suited for providing detailed and precise information about a particular case area (Jacobsen, 2017). Specifically, in this case, we aim to gain an in-depth understanding of the perspectives of both the designers and engineers. By identifying the differences between these two groups, we can then develop recommendations to improve collaboration between them.

Our theories are tested against interviews with engineers and designers for us to get different perspectives on the issue at hand. With the collected data from interviews, we aim to be able to see if there are any differences between the perspectives. Therefore the selection of the interviewees have determined the quality of the data, reflect the quality of the results and must be considered when conducting our questions, as well as interviewees.

Our decision to utilize semi-structured interviews was driven by the desire to incorporate both structured and unstructured questions in the interview process. This approach enabled us to gather primary data from both designers and engineers, facilitating a more objective comparison between the two groups (Jacobsen, 2017). To ensure more relevance of our data, we intentionally avoided using any other form of primary data collection, such as observations or target-group interviews since they did not align with our research objectives.

### 3.2 Data collection

With the gained knowledge, we were able to create our own theoretical framework (discussed in 3.3) which helps us to narrow down the scope of the interviews and prepare relevant questions within the problem area. The prepared interview questions have been asked to specific engineers and designers. Most of them are professors and lecturers at Swedish universities with experts in the fields of car intelligence and AI research, representing the engineer perspective, as well as professors specialized in informatics and information science, representing the designer perspective. In addition to academic professionals, we have also interviewed candidates working in the automobile industry, who have previously worked on AI-specific projects. Meaning they are not specifically specialized in AI, but at least have worked as a designer or developer in atleast one project including AI technology in the automobile industry.

The interviews that we prepared are going to be semi-structured interviews which means that we have predetermined questions within a specific theme, but the questions are open to start a discussion and we may not choose to use all questions and might ask questions generated on the spot (Jacobsen, 2017). The interviews are an ongoing process that have followed the whole project. Our goal is to have a fairly equal representation from both perspectives, e.g., three designers and three engineers to help us ensure non-biased towards any perspective, and provide more holistic and well-rounded analytics.

Throughout the duration of this study, we have engaged in a total of six interviews with individuals possessing a relevant background. These interviews were conducted remotely via online meetings with an average duration of 28 minutes each. The interviews are conducted online in order to transcribe what has been said. The questions that we have conducted strive from our theoretical framework where the questions are divided into four different categories which are (1) general questions about AI and ethics, (2) questions involving the trust with autonomous cars, (3) safety and (4) transparency. We have chosen to start with some general questions about the problem area to get a good discussion going and after that involve the different categories to get a better understanding of the interviewee´s perspective and thoughts surrounding the topic.

### 3.3 Theoretical framework

There is a correlation between trustworthiness and transparency, as transparency is a contributing factor to understanding the system, which results in trustworthiness. Furthermore, there's a correlation between reliability and safety, where a reliable technology can enhance the safety of the system. In addition, there is an alignment between all these factors (transparency, safety and reliability) that enables for a trustworthy system (see Figure 1).

The reason for the conclusion that reliability, safety, and transparency are correlated is based on the notion that these three factors contribute to creating a trustworthy AI system that can be trusted by users, stakeholders, and society as a whole. Trust is a key element in the development and adoption of autonomous vehicles, and it is a necessary factor for a successful deployment of these vehicles on a large scale and ultimately leading to increased trust and acceptance of autonomous vehicles in society. There are of course other factors that contribute to a trustworthy AI system, such as responsibility & accountability and privacy, however reliability, safety and transparency are the biggest challenges as well as the biggest contributors to trust as per analysis of the literature (Shneiderman, 2020a; Hjetland, 2015;

Wärnestål, 2021; Shneiderman, 2020b; Barredo Arrieta et al., 2020; Mikalef et al., 2022; Maurer et al., 2016; Ozmen et al., 2023; Charisi et al., 2017; Hengstler et al., 2016)

To reinforce the idea that trust is primarily fostered through transparency, safety, and reliability, Rothenberger et al. (2019) conducted a study to identify the most critical aspects of AI, including transparency and reliability, which were ranked as one of the most significant ethical considerations in the development of autonomous vehicles. While safety was not explicitly examined in this study, it is closely linked to reliability and should also be a top priority.

In the context of Figure 1, *transparency* can, as mentioned in the introduction, be divided into two parts: transparency for the user and transparency for the developer. When referring to the alignment triangle, transparency for the user is of greater importance as it is the user's trust that we strive to achieve by displaying the underlying reasoning for the decisions made by the system (Barredo Arrieta et al., 2020). *Reliability* pertains to the dependability of the system, including factors such as accuracy, downtime, and ensuring the absence of bugs. Reliability can be referred to as the consistent performance of a system as intended, with a specific emphasis on ensuring human accountability, fairness, and comprehensibility (Shneiderman, 2020b). *Safety* is a critical aspect of any system, as it affects not only the user but also the environment, other species, machines, and humans that could be affected by the system. A safe system is one that has been designed with these factors in mind and takes measures to prevent unintended harm and unexpected consequences. AI systems should be developed with a proactive risk prevention approach, ensuring that they behave as intended (Mikalef et al., 2022).

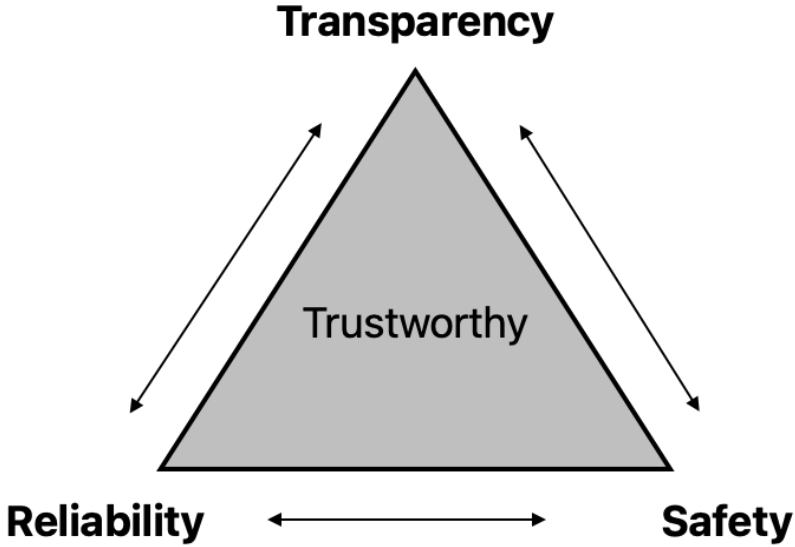

*Figure 1: Alignment for trustworthy AI.*

Our approach to data collection will be guided by our own created framework. The framework was created to verify the alignment of the three pillars in creating a trustworthy AI. We developed interview questions that are based on these three pillars, ensuring that our data collection process is consistent, systematic, and targeted towards obtaining information that is relevant to our research objectives. By focusing on these key pillars, we aim to ensure that the data we collect is informative and comprehensive, providing us with a deeper understanding of the factors that contribute to building trustworthy AI systems.

## 3.4 Analysis of collected data

The collected data from interviews were divided into two distinct groups: engineers and designers. This categorization was pre-established based on the participants' professional background or current occupation. It is important to note that the participants were not asked their preference for group assignment, and the categorization was solely based on our evaluation after the first question in the interview "Tell me about yourself, and what do you work with". Once the grouping was complete, we conducted a thorough analysis of the information to highlight any notable differences or outliers that may exist between the two groups.

In order to provide a more detailed analysis of the data, we further categorized the differences we found between the two groups into subfields that were deemed relevant. These subfields included safety, transparency, and reliability, among others. By breaking down the data into these subfields, we were able to gain a better understanding of how the differences between the two groups may impact each specific area. If we found that the differences were not applicable to any specific subfield, we used them as general differences.

Our approach focuses on identifying the most important and valuable information from qualitative data. To guide our analysis of the interviews, we are following the four-step process described by Jacobsen (2017) as a rule of thumb for qualitative data analysis.

*Document*: This step involves describing the material obtained through interviews.

*Explore*: Here, relationships that emerge from the data are identified through relatively unsystematic searching.

*Systematize and categorize*: The overwhelming information in qualitative data is reduced by dividing it into categories.

*Connect the dots*: In this step, the data is interpreted to make sense of the results.

## 3.5 Ethical considerations

With targeted interviews we need to look over ethical considerations of how we should present the information gathered. With the consent from the interviewee we will collect data that will be presented in this project. If the participant has any specific requests, such as not wanting the interview to be recorded, we need to respect this request and adapt our method accordingly. Before the interview, the interview questions were sent out in order for them to prepare or be well aware about the topics that will be discussed.

Consent is one of our highest priority related to data collection from interviews, therefore we want to ensure an informed consent. In order to achieve a fully informed consent, we follow 4 guidelines mentioned by Jacobsen (2017):

1. Interviewee must be able to make voluntary choices, ability to evaluate pros and cons.
2. No pressure from external person
3. Full information about the purpose of the interview and how it will be carried out
4. It must not only be informed about the study, it should be ensured that the interviewee actually understands it.

In addition to these considerations, confidentiality, privacy and respect should be considered. It's important to respect the interviewees boundaries and not push them to disclose anything they're uncomfortable with while ensuring confidentiality by keeping their personal information private.

Lastly, it's important for us to avoid bias in our interview questions in order not to lead the participant to certain answers that we are looking for. This will help us get a better view of the experience that the participant has and their own thoughts on the problem area. By not having our own bias with the questions we enable more thorough research. This was ensured by asking all participants the same line of questions that were objective and not biased towards any of the perspectives, and also asking open ended questions with the interviewee leading the conversation.

*Table 1: Description of Interviewees*

| Interviewee | Introduction |
|---|---|
| **Designer A** | Experience with the human-centric approach in designing, implementing, and evaluating digital innovations. Understanding people's needs in the context of digitization and strive to create valuable digital innovations that cater to those needs. |
| **Designer B** | Expertise in effects of IT on individuals, organizations, and society, with a particular focus on digital transformation in rural areas. I also explore e-government/e-governance and improving access to cultural materials. |
| **Designer C** | Working with user experience at leading transportation company, to streamline the use of vehicles. Have practical experience with UI/UX for AI systems. |
| **Engineer A** | Expertise in Cyber Physical Systems (CPS). Background in developing active systems for trucks, cars, and race cars, along with experience in connecting mechanical and electronic components. |
| **Engineer B** | Project manager: Digital Services and Systems, Embedded Intelligent Systems |
| **Engineer C** | Head of research: focuses on the increasing significance of the context in which AI systems operate |

# 4.0 Results & Analysis

In enlightening of an ongoing shift in the field of AI and Industry 5.0 (Leng et al., 2022; Xu et al., 2021; Shneiderman, 2020; Wärnestål, 2021; Shneiderman, 2020a; Shneiderman, 2020b) we assumed that people that have worked in the industry for a longer period of time has a less holistic view in regards to ethical dilemmas as compared to newly graduated engineers within the field. The reason for our assumption is due to the fact that there is an ongoing shift within the field of AI, and that this shift is researched and studied within universities and academic people. On the other hand, those who work in the industry may not always be aware of the changes taking place in their environment. While we were not able to confirm our theory regarding newly graduates having a more holistic view, we were able to see that engineers are more prone to the rationalistic view and tend to follow regulations or frameworks, such as EU legislation. Whereas designers follow user-centered approaches and overall have a more holistic view of the ethical and external implications of poorly designed AI systems.

We also draw a parallel between the ongoing shift with industry 5.0 and AI where both previously were defined with cutting-edge technologies that had high promise with impressive improvements to effectiveness, efficiency, and economical benefits. But not until recent times is it adding a new dimension; having a system/holistic view by having external and ethical considerations when developing the product or service.

Our analysis revealed that there are notable differences in how participants viewed the importance of transparency, reliability, and safety. While some participants prioritize transparency as the most crucial pillar, others place more emphasis on reliability or safety. Additionally, we found that some participants view these pillars as interdependent, whereas others see them as separate and distinct concepts. The results of our analysis are presented in separate chapters, each focused on a specific subfield related to transparency, reliability, and safety.

Discussions during the interview have brought attention to a shift in the distribution of responsibility for ethical considerations. In the past, car manufacturers were primarily responsible for providing a product that could transport individuals from one point to another (Mahut et al., 2017), with some ethical considerations around safety and environmental impact. However, in the age of connected and autonomous vehicles, car manufacturers now provide a service that includes transportation and various smart features such as being connected to the internet (Mahut et al., 2017), placing more emphasis on service delivery rather than just product production. Consequently, developers now bear greater responsibility for ensuring that their products and services are ethical, not only in terms of safety and environmental impact, but also in terms of privacy, security, transparency, and reliability and trust. These factors may vary across different regions of the world, adding to the complexity of the developers' responsibility to ensure ethical considerations.

Discussion with the respondents where focused on the fourth level of autonomy. Autonomy level four means that the car has high automation, the car can do almost everything by itself but needs human intervention under specific areas or scenarios, which can be weather, dirt roads etc.

## 4.1 Transparency

Transparency is a complex subject and one that we noted as most divergent in perspectives. One noticeable difference is the importance of visibility for the user of the underlying decision. While engineers tend to neglect this and focus only on transparency for the developer, stating that providing the underlying reasoning for the decision made could be overwhelming and distracting for the user. In contrast, designers note that transparency for the user is one of the most critical factors in building trust when using an AI system.

This is particularly noticeable when discussing the importance of transparency for end users and the challenge of balancing end user transparency with developer transparency. A typical example are the following statements from our interviews:

According to Engineer A, "*The vehicle needs to be certified. The end user does not need to know how it works*" (A. Engineer, personal communication, March 13, 2023).

Engineer B also stated, "*I mean if the car is driving itself and I'm simply sitting along as a passenger. I don't really need any information. What do I use it for? So, I think there already is probably enough information presented in the car when you sit and drive yourself, for example it starts pinging and it starts a lot of disturbing moments that I'm not really interested in. I just want the car to go forward. Yes, sometimes you don't know what it means. Some symbol comes up and pling pling pling and that's what I need done now*" (B. Engineer, personal communication, March 17, 2023, translated from Swedish).

Designer A raised concerns about transparency, stating, "*What kind of computer do I have, what kind of data does my car produce and it's quite miserable that I, as the owner of an artifact, i.e. the car, have no control over it. What are they finding out about my artifact that I own and pay for every month, and think they have the right to take data without informing me of what data they are taking*" (A. Designer, personal communication, March 30, 2023, translated from Swedish).

Designer B highlighted the importance of transparency, stating, "*It is clear that one must be transparent. If we now take autonomous cars as an example there, I think that a lot of people still don't understand the technology*" (B. Designer, personal communication, March 14, 2023, translated from Swedish).

Striking a balance between these perspectives is crucial. While it is agreed that providing information about every single decision and its underlying reasoning can be distracting, there are scenarios where the user does not agree with the decision made or where it results in an undesirable outcome, such as a collision. In such cases, transparency for the user becomes much more desirable and necessary to help them avoid ending up in the same situation again.

One critical finding from our study is that the two different types of transparency (for the user and for the developers) are often developed independently. This creates a gap between the designers and engineers, which can lead to a lack of coordination in the infrastructure and uneven development of the two different aspects of transparency. This is what we often see in the current autonomous cars where the transparency for the developer is more thought out than the transparency for the user. With designers unaware of how transparency for developers works and interacts with the system, and vice versa. This lack of collaboration can lead to conflicts and inconsistencies in transparency. To address this issue, we propose that both types of transparency should be developed, designed, and ideated together. This approach ensures that both developers and designers have a clear understanding of the decision-making process and underlying reasoning while ensuring that the information presented to the driver is not overwhelming or interferes with the transparency for the developer.

Both engineers and designers agree that if presenting information to the end user while driving, it must be clear, understandable, and not too complex. As Engineer C stated, "*So in the end, transparency doesn't mean that you show the code, but that you can explain why things happen*" (C. Engineer, personal communication, April 3, 2023). Designer A also highlighted the importance of presenting information in a user-friendly way, stating, "*What kind of information do I push back to the user and in what form because you can't claim that diagrams or statistics or algorithms and things like this are optimal because the users don't understand that"* (A. Designer, personal communication, March 30, 2023, translated from Swedish).

This indicates that designers offer a valuable perspective on this pillar, as they prioritize the user's needs and building trust. Designers tend to have a more holistic view of the matter and can provide crucial insights on how to ensure trust while still creating an effective and user-friendly AI system. However, this is not to say that one perspective is better than the other. But we argue that there already is a good level of transparency for developers, and it is now crucial to involve the user in the transparency process.

**4.2 Safety**

Safety can be arguably the most critical factor from an end-user perspective, and it is the area in our theoretical framework where we see the least difference in perspectives, which is not surprising. Despite coming from different backgrounds and having different viewpoints, both designers and engineers have good intentions and are committed to advancing this innovative field.

At its core, safety involves developing a system with the best possible intentions, and no one wants to see anyone get hurt, cause any harm, or be held responsible for something that is unsafe. Therefore, we observe that both perspectives are aligned with each other in their efforts to ensure that the system is developed and trained in a way that causes as little harm as possible to people, animals, and the environment.

We observe that the finding that safety is the most aligned view is also reflected in real-world examples such as Tesla's annual "Tesla Vehicle Safety Report," which provides data on the number of miles driven by their vehicles before encountering accidents (Tesla Vehicle Safety Report., n.d.). The report indicates that the average driver in the United States experiences an accident every 0.5 million miles driven, while the Tesla autopilot encounters an accident after an average of 5 million miles, making it ten times safer than the average American driver.

**4.3 Reliability**

To rely on autonomous cars is not effortless but we found that the engineers have more trust in technology and the performance of tests is an important factor in making the autonomous car reliable. The engineers are more likely to believe in the technology after the relevant tests have been done, no matter the state of the development of the car (beta testing). The engineers are more prone to early adopt new technologies to either try them out or to use them, we found out that engineers place their trust in the technology from the start, as long as it's been tested from the factory.

While the designers do not fully consider the autonomous car to be reliable until it has been user-tested by a larger percentage. In some cases the designers were more reliable towards the autonomous car if there was a bigger company that they were aware of. For example if the manufacturer was Volvo the designers were more prone to rely on the systems and the car before the larger number of people had tested it. Which Designer C describes, additional evidence is needed that the autonomous vehicle functions as intended and can be trusted. *¨So the driver has a track record of course, but then it would be linked, preferably to that context. Well, in that context one year, think about it. I had maybe a little more if you are going to have self-driving cars out on the road, maybe not just jump in and trust a car¨* - (C. Designer, personal communication, March 20, 2023, translated from Swedish). According to designer C, autonomous cars are more reliable after approximately one year of driving in society. With the approach that autonomous cars need more testing before it is reliable, which is common from the designer perspective.

Trusting autonomous cars is a big issue in the development of AI. With our results, two different viewpoints were presented on how the two groups viewed an autonomous vehicle as being reliable. Engineer A suggests that the major car companies use a variety of tools to ensure that the car is fully functional and can be trusted. *¨The automotive industry has testing bodies and the press they will flag anything that is not correct.¨* (A. Engineer, personal communication, March 13, 2023)

During an interview, Designer B expressed skepticism about riding in an autonomous car, highlighting concerns about safety and the absence of a human driver. As they stated, "*I think it would be very difficult for me to sit in an autonomous car and think OK, I'm just going along with you and me. I see no driver, I don't see anyone who can brake or swerve or similar. For me then it would be far ahead of me*" (B. Designer, personal communication, March 14, 2023, translated from Swedish).

This skepticism about trust in autonomous cars is shared by Designer A and C, who also stressed the importance of testing and adoption before accepting these technologies. With this in mind we see that designers tend to have a harder time relying on the system and may need more time to see more widespread adoption before using the service themselves.

On the other hand, Engineer A suggests that regulatory measures could increase trust in autonomous vehicles, stating the following after receiving a question related to if trust can be achieved solely by impressive technology, "*You have to have regulations, measures. So the moment you get traffic regulations that include autonomous vehicles, you get more trust that things actually work*" (A. Engineer, personal communication, March 13, 2023). This statement suggests that engineers, being early adopters of new technology, have more trust in the capabilities of autonomous vehicles. However, they believe that the current laws and regulations have not kept up with the advancements in technology. On the other hand, designers are more skeptical and cautious in their approach towards autonomous vehicles.

Literature specialized in design and design thinking such as Lewrick et al (2018), encourage designers to be innovative, explorative, and bold by nature. They use design thinking to solve wicked problems with undefined outcomes. In contrast, engineers are typically more focused on problem-solving, technical analysis, and try to solve the issue by developing or redesigning technology. They respect the potential of technology, and understand it better.

When it comes to AI, being innovative and explorative, like a designer, can be valuable, but it may cause us to move too quickly and want to achieve mass adoption too rapidly. We need to remember the importance of respecting technology and making it reliable before we continue to innovate further. In the case of AI, where the decisions can be lethal, we believe it's crucial to prioritize reliability over innovation in some cases, such as reliability to achieve trust.

As we develop autonomous cars, we may encounter wicked problems that arise as unintended consequences. A designer's mindset can be valuable in addressing these challenges such as environmental impact, ethics, and other complex issues that arise from the development of such technology.

To build reliable AI, we need a problem-solving and tech-forward approach that prioritizes reliability to gain public trust. The reliability issue is primarily related to the technology itself, so we must adopt a mindset that values technical proficiency and analytical rigor to ensure the safe and reliable operation of autonomous vehicles.

### 4.4 Analysis of the results

The result that we have obtained reflects differences in relation to reliability, which is described by Shneiderman (2020b) as an important part of creating trust for its users. The differences presented in the result is the approach the two groups have to what they themselves consider important to gaining a reliable system. Engineers tend to place great trust in a system's reliability when it has undergone thorough testing and has a track record of success. The results of well-executed testing provide engineers with proof of a system's reliability, which reinforces their trust in it. In contrast, designers are more likely to place emphasis on transparency and widespread adoption. While these factors may indicate a system's popularity or usability, they do not necessarily guarantee reliability.

As mentioned in the introduction, transparency can be divided into two distinct areas - user-based and developer-based (Barredo Arrieta et al., 2020). Through interviews with industry professionals, we have confirmed that these two types of transparency are independent of each other and are typically not developed together. Our findings suggest that perspectives on transparency are most aligned within the same group and differ significantly across the two different groups. We argue for a more inclusive approach to transparency that prioritizes the end-user's needs. Engineers should move away from solely focusing on developer transparency, as it is ultimately the engineer who will develop both types of transparency. By prioritizing both user and developer transparency, we can improve the credibility and overall success of the product since transparency is one of the most important factors for credibility (Barreto Arrieta 2020) .

Out of all the tree pillars, safety was the one that both the groups had similar thoughts about. In Figure 2 both designers and engineers classified safety as high importance when developing AI for autonomous cars. Safety is an important factor in the development of autonomous cars and AI systems, which the results clarified by the equal importance from both perspectives.

Hengstler et al., (2016) presents that one important factor to achieving trust is through the involvement and consideration of safety and ethical perspective. The study by Hengstler et al. (2016) concludes that while safety is crucial for trust, it alone is insufficient to establish complete trust from users. As a result, the study suggests the exploration of new subfields that can help in building user trust in AI. Therefore, if all these three pillars are not carefully implemented or considered in the implementation of autonomous cars, the end user will never achieve the last dimension of faith or overcome perceived risk (Hengstler et al., 2016). Overcoming perceived risk is why establishing trust in autonomous driving should be a top priority in order to ensure that users ultimately rely on the technology. However, developing trustworthy systems can present significant challenges, especially when there are differing views and ethical frameworks that must be understood that often are vague and lack concrete guidance for developers and users of AI (Hengstler et al., 2016). In addition, it is important to determine the role of humans in relation to the system and navigate these challenges in order to build the necessary trust in autonomous driving.

Additionally, our research has shown that in order to achieve the tasks we aim to accomplish with AI and to achieve the future vision of our respondents, it is necessary to grant AI partial authority and a certain level of control over the user. This finding aligns with the theories of Hengstler et al. (2016) regarding the interplay between human and machine control. However it is equally important to allow the user to retain some degree of agency and control (Shneiderman., 2020a; Shneiderman., 2020b;). Both designers and engineers agreed that control for the end user still is important, but AI should still be able to take control over the task and perform it faster and more accurately than the human. In other words, effective collaboration between humans and machines requires a delicate balance between automation and human input.

To achieve this balance, it is crucial to carefully define the appropriate level of automation for a given task (Hengstler et al., 2016). This requires taking into account factors such as the complexity of the task, the capabilities of the AI system, and the preferences and abilities of the human user. By finding the right level of automation, we can create a harmonious partnership between humans and machines that allows us to achieve greater efficiency and effectiveness in our work, instead of being fully replaced by machines.

**4.5 Differences between designers and engineers**

In Figure 2, we present a comparison of the perspectives of designers and engineers who were interviewed on the three fundamental aspects of trust: *transparency, safety, and reliability.* While there are some areas where both groups share similar viewpoints, such as safety, there are also notable differences in their opinions. For instance, designers tend to place more emphasis on transparency as a means of building trust, whereas engineers prioritize reliability. Nevertheless, both groups acknowledge the significance of security in the development of autonomous cars.

It's important to note that Figure 2 illustrates the varying levels of knowledge and importance of designers and engineers in different fields of AI, with the aim of achieving a trustworthy AI. The triangles in the Figure represent each group and highlight the areas where the differences between them are the greatest. To facilitate comprehension of the Figure, we have established interwalls consisting of values ranging from 0 to 1. The values indicate the level of importance assigned to each field by both designers and engineers. To clarify the parameter, we have set the range of 0.8 - 1.0 as high importance, 0.4 - 0.79 as medium importance, and 0 - 0.39 as low importance for both groups. Being positioned in the 0.8 - 1.0 indicates that their perspective is better understood by the end-users and has a greater potential to achieve trust. For instance, the emphasis on transparency by designers means that they are more likely to involve users in the transparency process, which is an approach we encourage. Conversely, engineers have a better understanding of the underlying technology and can determine when the software is reliable enough to support innovation, which also is an approach we encourage.

It's important to emphasize that effective communication between designers and engineers is crucial to achieve transparency and reliability in the development of autonomous cars. As we mentioned earlier, transparency should be designed together, with the end-user involved in the process to determine which information should be visible. This ensures that the end-user can understand the technology, its limitations, and its potential, which is essential in building trust.

Therefore, it's important for designers and engineers to collaborate and communicate effectively to develop a transparent and reliable framework for autonomous cars. By working together, they can leverage their respective strengths to create a comprehensive and trustworthy system that meets the needs and expectations of end-users.

The interviewed respondents were asked if they considered humans to be the weakest link in the ecosystem in relation to self-driving cars. The respondents from both the designer and engineering profession were very united on that issue. Although most participants in our study agreed that humans were the weakest link of AI systems, there was one notable outlier. This participants argued that if an AI system is unable to react or collaborate effectively with human drivers in traffic, then the system may actually be the weakest link.

This insight highlights the importance of considering the interaction between AI systems and their human users when evaluating the effectiveness and reliability of these systems. While it's true that humans may introduce errors or biases and not be able to do tasks as efficiently or precisely, it's also important to recognize that effective collaboration between humans and AI is essential for ensuring the safe and efficient deployment of these systems.

In addition, we found that both engineers and designers often disagree on how ethics are implemented in their projects. They typically rely on existing guidelines and frameworks from the EU legislation and the project's customer to resolve these issues. Although both groups believe that there are already enough frameworks and legislation in place, they do agree that these guidelines should be taught as part of their education, either in school, or at the company since the projects tend to disregard these guidelines or not have a full understanding of them. By incorporating ethics into their education, team members can better understand the existing frameworks and legislation, which can help ensure that they have a shared goal of developing trustworthy AI for autonomous cars. Early education in ethics would emphasize its importance and enable individuals to create products that meet customer expectations while incorporating ethical considerations.

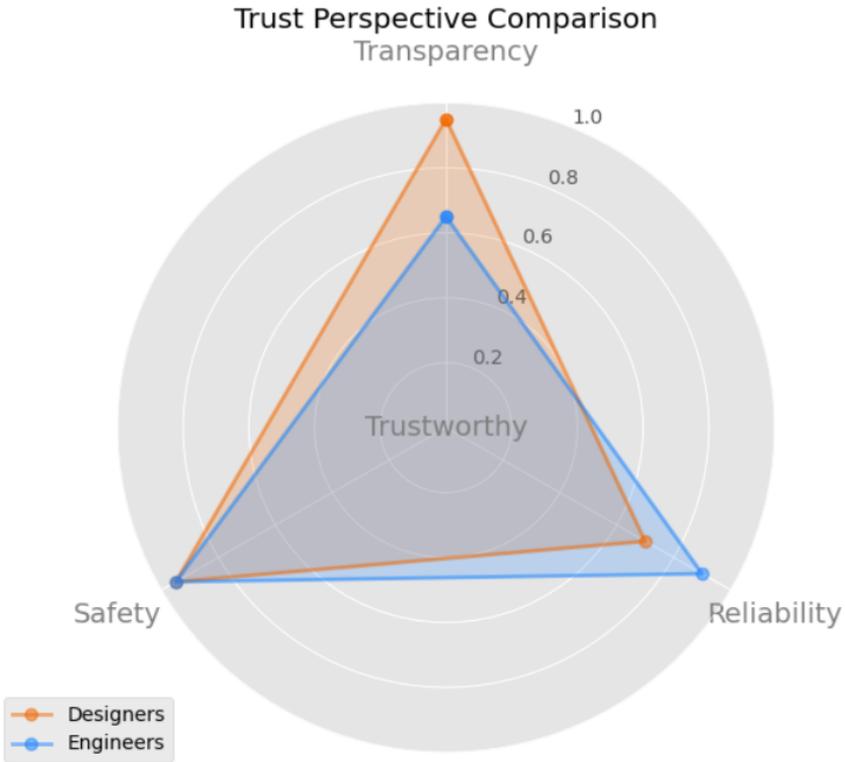

Figure 2, Perspective Comparison

While collecting data, we received various insights on how ethical considerations could be improved in AI-related projects for autonomous cars. These insights are summarized in Table 1. By utilizing these findings, we were able to establish recommendations for both designers and engineers to incorporate ethical considerations into the development process.

*Table 2, summary of differences and recommendations that can be used as strategies*

| **Subfield of ethics** | **Differences** | **Recommendation** |
|---|---|---|
| **Transparency** | Engineers prioritize access to technical attributes and information about the car's decision-making process, while designers are primarily concerned with presenting this information in a digestible and user-friendly format that is accessible to non-technical individuals. The engineers do not believe that this information is crucial for the driver, while the designers are not particularly interested in the underlying technical aspects of the decision-making process. | Engineers and designers should collaborate throughout the development of transparency features in autonomous cars. Including end-users in the process can ensure that transparency is developed with all stakeholders in mind, and that the technology is not limited in the future due to any specific considerations made during the development process |
| **Reliability** | Engineers tend to have more trust in the technology of autonomous cars, especially after relevant tests have been done. Designers, on the other hand, are more likely to view the reliability of autonomous cars with greater caution, preferring to wait until a larger percentage of users have tested the technology before fully trusting it. | It is essential that designers are well-informed about autonomous vehicle technology and are actively involved in the testing processes. This would increase their knowledge of the technology and create more trust. Furthermore, experts and professionals working in this field should take a proactive role in shaping regulations and laws related to AI technologies. |

| Safety | Both designers and engineers are committed to advancing the field with good intentions. The perspective of safety is aligned in both fields and is also reflective in the use-case of autonomous cars where Tesla autopilot is ten times safer than the average American driver. | Although their perspectives are aligned towards safety, it is crucial for designers and engineers to remain up-to-date with the latest advancements to ensure that the technology continues to progress and become even safer and make noticeable adjustments if needed, such as bad reports from the Tesla autopilot. |
|---|---|---|

# 5.0 Discussion

## 5.1 Insights & Reflections

Effective communication is a critical aspect of any project, particularly when it comes to developing AI systems. As these systems continue to be ubiquitous in our daily lives, it's essential to prioritize the needs of end-users. AI is intended to make our lives easier and more convenient, and to achieve that goal, we must prioritize communication and collaboration in the development and deployment of these systems. In this thesis project, we explored the importance of adopting a united view when creating trustworthy AI. Our study results highlight the pivotal role of communication and end-user involvement in ensuring successful AI deployment.

While our study's reliability may be affected by the sample size of our interviews, it's important to note that, toward the end of our interview process, we began to receive similar responses from participants with similar backgrounds. Despite this limitation, we believe that the insights gained from the interviews provide sufficient information to make generalizations about the study's findings. We believe that in order to start receiving more valuable responses, the sample population needs to increase and involve more participants from different companies within the industry. With gained insights from people working on projects where designers and engineers collaborate together and use different guidelines would be valuable to get a better understanding of the differences.

During our interviews, we posed a thought-provoking question to our participants: Should cars be developed to have emotions? Specifically, should a car's speed and driving behavior be adapted to the driver's emotions, such as driving faster when the driver is feeling stressed? While everyone we asked agreed that a car should not drive faster because the driver is late to work or feeling stressed, the question became more relevant in situations where the driver is elderly and unable to handle a car's speed or in life-threatening situations where getting to the hospital quickly is the top priority. The reason this question was not mentioned in the results is because it is more philosophical in nature. We did not observe any patterns where designers or engineers had divergent views. Rather, it seemed to come down to personal traits and whether they had elderly family members. While this is likely an easy function to incorporate into a car, we found this subject interesting to discuss with various individuals due to the varying opinions.

It is crucial to acknowledge that AI systems are not inherently responsible for their actions. Therefore, in the event of a system failure, the blame cannot be solely attributed to the AI system itself. Responsibility lies with all parties involved in the development, implementation, and operation of the system. These systems are simply tools created and used by humans, which raises significant ethical questions about their role in society and the obligations of those who develop and deploy them. While this topic goes beyond the scope of our research, it is critical to consider in discussions about the trustworthiness of AI systems.

Trust is a grand issue and, in order to have trust for an AI system, the user needs to be informed and fully aware of its capabilities. The future of AI systems and how they will be integrated into our society is a subject that often receives conflicting views. Some people are overly optimistic about the potential benefits, while others are overly pessimistic about the risks. It's important to pay close attention to building the appropriate level of trust in AI systems (Charisi., 2017).

To conclude, the topic of AI is complex and multifaceted. While we have explored some of the issues and debates surrounding the ethics of it, there is still much more to learn and discover. As researchers continue to delve deeper into this field, we may gain a better understanding of its nuances and implications for society. Whether you are an expert in the field or simply curious about it, AI, and specifically autonomous vehicles is an area that offers endless opportunities for exploration and learning and in most scenarios in AI ethics, there is never a right or wrong.

**5.2 Putting Recommendations into Practice: Approaches and Considerations**

Based on the recommendations derived from the analysis, the next step is to explore practical approaches for implementing them. This involves going beyond the high-level recommendations to identify specific strategies that can be put into practice. By developing a comprehensive approach to implementing these recommendations, organizations can more effectively address ethical concerns related to AI in autonomous vehicles. In order for this to work we have come up with two different ways to bring together the groups to influence and educate them and achieve a trustworthy system.

To effectively implement the recommendations of the framework, workshops involving stakeholders from different backgrounds are recommended to create a shared understanding of ethical principles and embed them in the development process. Cross-functional teams, consisting of designers, engineers, ethicists, and other relevant stakeholders, can also work together to ensure that ethical considerations are integrated into the development process. To ensure effectiveness over time, regularly reviewing and updating the implementation strategies is crucial. Additionally, measuring the impact of the strategies can be done by having cross-functional teams conduct a workshop and comparing their perspectives with those of a team that did not undergo the workshop.

In addition to workshops, another approach to implementing the recommendations is to establish cross-functional teams consisting of designers, engineers, ethicists, and other relevant stakeholders. These teams can work together to ensure that ethical considerations are integrated into the development process from the very beginning. It is important to establish clear communication channels and protocols within these teams to ensure that all perspectives are heard and considered. Additionally, regularly reviewing and updating the implementation strategies can ensure that the recommendations remain relevant and effective over time. It is also important to note that this should be incorporated into the development process at an early stage, where both designers and engineers can learn from each other and get a more holistic understanding of the ethical considerations that come from both sides.

Lastly, we recommend being involved in as many processes as possible, even though it might not be directly linked to your working tasks. By involving designers in testing processes and troubleshooting for the technology, they gain a deeper understanding of autonomous vehicle technology and become actively engaged in the testing procedures. This not only enhances their knowledge of the technology but also promotes effective collaboration between engineers and designers. In addition designers, engineers and end users should collaborate in the processes of developing and AI features that need any transparency visualization for the end user.

While these approaches can be applicable to most companies, measuring the success of their implementation can pose challenges. However, implementing these approaches allows for the evaluation of trust improvement over time through case studies and questionnaires. These tools can capture changes in participants' responses and findings throughout the implementation process, providing valuable data for analysis.

One method for measuring participants' ethical decision-making abilities is the use of case studies and scenarios. By presenting real-life ethical dilemmas or hypothetical situations, we can gauge their capacity to analyze and respond to ethical challenges. Evaluating their responses and decisions and how they changed over time can serve as an indicator of their progress in applying ethical principles and reasoning.

Discussion is an alternative measurement in order to foster deeper understanding and collaboration. Involving participants in discussions about ethical issues opens up for critical thinking and exploration of different perspectives. Having someone able to monitor and evaluate the discussions enables the participants for a deeper ethical understanding which can be used to evaluate the participants progress in ethical development.

# 6.0 Conclusion

The research questions for this thesis were as follows:

1. *What is the difference in perspective between designers and engineers with regard to ethics when developing autonomous vehicles?*

2. *What strategies can be implemented to bridge the gap in perspectives between designers and engineers regarding the development of autonomous cars?*

To answer these research questions, our research has found that there are differences in perspectives between designers and engineers when it comes to ethics in developing autonomous vehicles. While both groups acknowledge the importance of safety and security, designers tend to prioritize transparency, while engineers prioritize reliability. However, effective communication and collaboration between designers and engineers can lead to the development of a comprehensive and trustworthy framework for autonomous vehicles that meets the needs and expectations of end-users. Our research also highlights the importance of considering the interaction between humans and AI in evaluating the effectiveness and reliability of autonomous vehicle systems. Additionally, we recommend ethics to be incorporated into education of designers and engineers to ensure a shared understanding of existing frameworks and legislation and to enable the creation of products that meet customer expectations while incorporating ethical considerations. While it's important for developers and engineers to have education in regards to ethical AI, it should also be noted that if AI continues to develop in the same phase as it has in the last years, AI will have the potential to become as much of a necessity to our daily lives as the internet. Thus, we believe that education of ethical and responsible AI should be implemented in the school system at an early age e.g., elementary school to make sure that the country stays or becomes at the forefront of an evolving and impactful technology. The responsibility for ensuring ethical considerations are incorporated into AI should be shared among all parties involved, including universities, workplaces, and individuals working on AI projects. By doing so, a common ground for ethical practices can be established, making it an essential aspect of future AI development.

By reducing the gap in perspective between designers and engineers it is essential to develop a transparent and reliable framework for autonomous cars that meets the needs and expectations of end-users. Collaboration and effective communication between designers and engineers are critical to achieve transparency and reliability in the development of autonomous cars. It's important to involve the end-user in the process of determining which information should be visible and to consider the interaction between AI systems and their human users when evaluating the effectiveness and reliability of these systems. Furthermore, incorporating ethics into the education of designers and engineers can help ensure that they have a shared goal of developing trustworthy AI for autonomous cars. By leveraging their respective strengths and working together, designers and engineers can create a comprehensive and trustworthy system for autonomous cars.

While our findings enabled for theoretical strategies to bridge any discrepancies in ethical views, these strategies have not been field tested, thus we recommend for future research to apply our recommendations in workplaces and real life projects in order to see if the gap is reduced to improve the trust importance of autonomous cars. In addition, we have noted that ethics is not limited to transparency, reliability and safety and researchers could explore other factors that influence trust in AI systems, such as security, integrity and responsibility. By analyzing the interplay between these factors and the recommendations proposed in this study, researchers could provide a more comprehensive understanding of how to build trust in AI systems. Ultimately, the goal would be to identify the most effective strategies for increasing trust in AI systems and to promote their widespread adoption in the development of future systems.

# 7.0 References

Auernhammer, J. (2020). Human-centered AI: The role of human-centered design research in the development of ai. *DRS2020: Synergy*. https://doi.org/10.21606/drs.2020.282

Barredo Arrieta, A., Díaz-Rodríguez, N., Del Ser, J., Bennetot, A., Tabik, S., Barbado, A., Garcia, S., Gil-Lopez, S., Molina, D., Benjamins, R., Chatila, R., & Herrera, F. (2020). Explainable artificial intelligence (XAI): Concepts, taxonomies, opportunities and challenges toward responsible AI. *Information Fusion*, *58*, 82–115. https://doi.org/10.1016/j.inffus.2019.12.012

Breque, M., De Nul, L., and Petridis, A., "*Industry 5.0: Towards a sustainable, human-centric and resilient European industry,*" Luxembourg, LU: European Commission, Directorate-General for Research and Innovation, 2021. https://op.europa.eu/en/publication-detail/-/publication/468a892a-5097-11eb-b59f-01aa75ed71a1/

Charisi, V., Dennis, L., Fisher, M., Lieck, R., Matthias, A., Slavkovik, M., ... & Yampolskiy, R. (2017). *Towards moral autonomous systems*. arXiv preprint arXiv:1703.04741. https://arxiv.org/pdf/1703.04741.pdf

Chonko, L. (2012). *Ethical theories*. Retrieved 29th March, 2023 from https://www.dsef.org/wp-content/uploads/2012/07/EthicalTheories.pdf

Hengstler, M., Enkel, E., & Duelli, S. (2016). Applied Artificial Intelligence and Trust—the case of autonomous vehicles and medical assistance devices. *Technological Forecasting and Social Change*, *105*, 105–120. https://doi.org/10.1016/j.techfore.2015.12.014

Hjetland, S. M. (2015). *Designing for Trust in Autonomous Systems*. Department of Product Design Norwegian University of Science and Technology.

Jacobsen, D. I. (2017). *Hur genomför man undersökningar?: Introduktion till Samhällsvetenskapliga Metoder*. Studentlitteratur.

Javed, S., Javed, S., van Deventer, J., Mokayed, H. and Delsing, J., 2023. A Smart Manufacturing Ecosystem for Industry 5.0 using Cloud-based Collaborative Learning at the Edge. In IEEE/IFIP Network Operations and Management Symposium 2023-MFI5.0.

Khan, M.A.U., Nazir, D., Pagani, A., Mokayed, H., Liwicki, M., Stricker, D. and Afzal, M.Z., 2022. A Comprehensive Survey of Depth Completion Approaches. Sensors, 22(18), p.6969.

Leng, J., Sha, W., Wang, B., Zheng, P., Zhuang, C., Liu, Q., Wuest, T., Mourtzis, D., & Wang, L. (2022). Industry 5.0: Prospect and retrospect. *Journal of Manufacturing Systems*, *65*, 279–295. https://doi.org/10.1016/j.jmsy.2022.09.017

Lewrick, M., Link, P., & Leifer, L. J. (2018). *The design thinking playbook: Mindful digital transformation of teams, products, services, businesses and Ecosystems*. Wiley.

Mahut, F., Daaboul, J., Bricogne, M., & Eynard, B. (2016). Product-service systems for servitization of the automotive industry: A literature review. *International Journal of Production Research*, *55*(7), 2102–2120. https://doi.org/10.1080/00207543.2016.1252864

Maurer, M., Gerdes, J. C., Lenz, B., & Winner, H. (2016). *Autonomous driving: technical, legal and social aspects*. Springer Nature.

Mikalef, P., Conboy, K., Lundström, J. E., & Popovič, A. (2022). Thinking responsibly about responsible AI and 'The dark side' of ai. *European Journal of Information Systems*, *31*(3), 257–268. https://doi.org/10.1080/0960085x.2022.2026621

Mokayed, H., Meng, L.K., Woon, H.H. and Sin, N.H., 2014. Car Plate Detection Engine Based on Conventional Edge Detection Technique. In The International Conference on Computer Graphics, Multimedia and Image Processing (CGMIP2014). The Society of Digital Information and Wireless Communication.

Mokayed, H., Quan, T.Z., Alkhaled, L. and Sivakumar, V., 2022, October. Real-Time Human Detection and Counting System Using Deep Learning Computer Vision Techniques. In Artificial Intelligence and Applications.

Mokayed, H., Nayebiastaneh, A., De, K., Sozos, S., Hagner, O. and Backe, B., 2023. Nordic Vehicle Dataset (NVD): Performance of vehicle detectors using newly captured NVD from UAV in different snowy weather conditions. arXiv preprint arXiv:2304.14466.

Ondruš, J., Kolla, E., Vertaľ, P., & Šarić, Ž. (2020). How do autonomous cars work?. *Transportation Research Procedia*, *44*, 226-233.

Ozmen Garibay, O., Winslow, B., Andolina, S., Antona, M., Bodenschatz, A., Coursaris, C., ... & Xu, W. (2023). Six Human-Centered Artificial Intelligence Grand Challenges. *International Journal of Human–Computer Interaction,* 1-47.

Rothenberger, L., Fabian, B., & Arunov, E. (2019). *Relevance of Ethical Guidelines for Artificial Intelligence-a Survey and Evaluation*. In *ECIS*.

Rödel, C., Stadler, S., Meschtscherjakov, A., & Tscheligi, M. (2014, September). Towards autonomous cars: The effect of autonomy levels on acceptance and user experience. In *Proceedings of the 6th international conference on automotive user interfaces and interactive vehicular applications* (pp. 1-8).

Shneiderman, B. (2020a). Human-centered artificial intelligence: Three fresh ideas. *AIS Transactions on Human-Computer Interaction*, 109–124. https://doi.org/10.17705/1thci.00131

Shneiderman, B. (2020b). Human-centered artificial intelligence: Reliable, safe & trustworthy. *International Journal of Human–Computer Interaction*, *36*(6), 495-504.

Shneiderman, B. (2022). Defining reliable, safe, and trustworthy systems. *Human-Centered AI*, 53–56. https://doi.org/10.1093/oso/9780192845290.003.0007

Tesla Vehicle Safety Report. (n.d.). Retrieved March 31, 2023, from https://www.tesla.com/ VehicleSafetyReport

Wärnestål, P. (2021). *Design av Ai-Drivna Tjänster*. Studentlitteratur AB.

Xu, X., Lu, Y., Vogel-Heuser, B., & Wang, L. (2021). Industry 4.0 and industry 5.0—inception, conception and perception. *Journal of Manufacturing Systems*, *61*, 530–535. https://doi.org/10.1016/j.jmsy.2021.10.006

Yampolskiy, R. V. (2020). On defining differences between intelligence and artificial intelligence. *Journal of Artificial General Intelligence*, *11*(2), 68-70.

# 8.0 Appendix A: Interview questions

**Interview questions:**

1. If you have, can you tell us about a time when you worked with a team of designers and engineers to develop a product? How did you navigate any disagreements or differences in opinions between the two groups (designers/engineers)? - Specifically ethical disputes if possible. And how did their responsibilities differ? - IF no - Question 2
2. What role do you think designers and engineers play in ensuring that autonomous vehicles (or AI products) are designed and developed in an ethical manner? How might their responsibilities differ?
3. How do you think collaboration between designers and engineers can help to ensure that ethical considerations are integrated for autonomous vehicles (or products)? What steps can be taken to facilitate this collaboration?
4. In your opinion, what are some of the biggest challenges facing the design and development of autonomous vehicles from an ethical perspective? How do you think these challenges could be addressed?
5. Do you think that humans are the weakest link in the ecosystem? Meaning, should we design AI to make human tasks more precise than humans, or should it be complementary to the human capabilities? Human in the center or the technology feasibility in focus?
6. Would it be easier to account for ethics if there were clear guidelines, or should the developers adapt using a human-centered approach?

**Trust related questions:**

7. What do you think is the main factor that could turn us over the tipping point where the majority of users can trust autonomous vehicles?
   a. Should the companies and cars have a proven record with high accuracy?
   b. Should the user be able to "test drive" and make up their own mind? Word of month? Own experiences? Other examples?
8. Do you think that trust can be gained solely by impressive technology and a good track record? if no, why not?

**Transparency related questions:**

9. Should the driver always be able to see why the car made the decision? If yes, then why?
   a. Or do you think the driver doesn't care as long as the vehicle transported the user from A to B in a safe, responsible and ethical manner?

**Safety related questions:**

10. What are some of the biggest security and safety risks associated with autonomous cars in your opinion?

**Specific questions for people working with AI:**

11. What do you think are some of the key ethical considerations that need to be taken into account when designing AI services? How might these considerations differ between designers and engineers?
12. Can you give an example of a situation where a design decision for an AI product might conflict with ethical considerations? How would you approach resolving this conflict?